\documentclass[preprint,12pt]{elsarticle}
\usepackage{amsfonts,amssymb,amsmath,graphicx,amscd,ulem,times}
\usepackage{array}
\usepackage[latin1]{inputenc}

\newcommand{\tauc}{\tau_y}
\newcommand{\etaHB}{\eta_{\text{\sc hb}}}
\newcommand{\vFHB}{\v{F}}
\newcommand{\FHB}{\text{F}}
\newcommand{\gammadot}{\dot \gamma}
\newcommand{\gammadotG}{{\dot \gamma}_{f}}
\newcommand{\Gammadot}{\dot \Gamma}
\renewcommand{\v}[1]{\underline{#1}{}}
\newcommand{\vv}[1]{\boldsymbol{#1}}
\newcommand{\pG}{p_{f}}
\newcommand{\ps}{p_{\text{p}}}
\newcommand{\Vs}{V}
\newcommand{\vs}{v_{\text{p}}}
\newcommand{\vn}{\v{n}}
\newcommand{\vu}{\v{u}}
\newcommand{\vuG}{\v{u}_{f}}
\newcommand{\vus}{\v{u}_{\text{p}}}

\newcommand{\ez}{\v{e}_z}
\newcommand{\dive}{\text{div}\:}
\newcommand{\cauchy}{\vv{\sigma}}
\newcommand{\cauchys}{\vv{\sigma}_{\text{p}}}
\newcommand{\vvtau}{\vv{\tau}}
\newcommand{\vvtaus}{\vv{\tau}_{\text{p}}}
\newcommand{\vvtauG}{\vv{\tau}_{f}}
\renewcommand{\d}{\vv{d}}
\newcommand{\dG}{\vv{d}_{f}}
\newcommand{\dun}{\vv{d}_{1}}
\newcommand{\ds}{\vv{d}_{\text{p}}}
\newcommand{\tun}{\vv{\delta}}

\def\bi{\begin{itemize}}

\def\ei{\end{itemize}}
\def\be{\begin{equation}}
\def\ee{\end{equation}}
\def\bea{\begin{eqnarray}}
\def\eea{\end{eqnarray}}

\def\dij{d_{ij}}
\def\drt{d_{r\theta}}

\def\drz{d_{rz}}
\def\tij{\tau_{ij}}

\def\trt{\tau_{r\theta}}

\def\trz{\tau_{rz}}

\def\gdot{\dot\gamma}

\def\phid{\phi_\text{div}}

\def\m1{$^{-1}$}

\DeclareTextSymbol{\degre}{OT1}{23}

\journal{Journal of Non-Newtonian Fluid Mechanics}

\begin{document}

\begin{frontmatter}



\title{Shear-induced sedimentation in yield stress fluids}


\author{Guillaume Ovarlez\corref{cor1}}
\author{François Bertrand, Philippe Coussot, Xavier Chateau}
\address{Université Paris-Est, Laboratoire Navier (UMR CNRS 8205), Champs-sur-Marne, France }
\cortext[cor1]{corresponding author: guillaume.ovarlez@ifsttar.fr}

\begin{abstract}
Stability of coarse particles against gravity is an important
issue in dense suspensions (fresh concrete, foodstuff, etc.). On
the one hand, it is known that they are stable at rest when the
interstitial paste has a high enough yield stress; on the other
hand, it is not yet possible to predict if a given material will
remain homogeneous during a flow. Using MRI techniques, we study
the time evolution of the particle volume fraction during the
flows in a Couette geometry of model density-mismatched
suspensions of noncolloidal particles in yield stress fluids. We
observe that shear induces sedimentation of the particles in all
systems, which are stable at rest. The sedimentation velocity is
observed to increase with increasing shear rate and particle
diameter, and to decrease with increasing yield stress of the
interstitial fluid. At low shear rate (`plastic regime'), we show
that this phenomenon can be modelled by considering that the
interstitial fluid behaves like a viscous fluid -- of viscosity
equal to the apparent viscosity of the sheared fluid -- in the
direction orthogonal to shear. The behavior at higher shear rates,
when viscous effects start to be important, is also discussed. We
finally study the dependence of the sedimentation velocity on the
particle volume fraction, and show that its modelling requires
estimating the local shear rate in the interstitial fluid.

\end{abstract}


\end{frontmatter}

\section{Introduction}\label{section_intro}

Dense suspensions arising in industrial processes (concrete
casting, drilling muds, foodstuff transport...) and natural
phenomena (debris-flows, lava flows...) often contain coarse
particles that tend to settle as they are denser than the average
system density. This is a critical problem: if settling occurs,
the materials may lose their homogeneity, which can strongly
affect their mechanical properties. In slow flows, when the solid
particles are immersed in a fluid, it is considered that the
settling properties of suspended particles are not significantly
affected by the material flow, and the sedimentation velocity is
usually computed from the balance of gravity and drag forces. In
order to avoid or slow down sedimentation, the only practical
solution consists in inducing a sufficient agitation to the system
which will induce some lift or dispersion forces to the particles.
This principle is typically used in fluidization process, in which
a vertical flow of the interstitial fluid induces a drag force
counterbalancing gravity force. For horizontal flows in conduits
one may also rely on turbulence effects
\cite{Oroskar1980,Davies1987} or on viscous resuspension
\cite{Leighton1986,Zhang1994}.

For many materials, the situation is different: the denser
particles do not settle at rest because they are embedded in a
yield stress fluid which is able to maintain the particles in
their position. This situation is typically encountered with
mortars or fresh concrete \cite{Roussel2006, Roussel2011} which
are made of particles (sand or gravel) of density around 2.5 mixed
with a cement-water paste of density around 1.5. This is the same
for toothpastes which contain silica particles of density 2.5
suspended in a paste of density close to 1. Basically, the net
gravity force exerted on particles of diameter $d$ suspended in a
yield stress fluid of yield stress $\tau_y$ is counterbalanced by
the elastic force exerted by the interstitial material as long as
$\Delta\rho g d \lesssim\tau_y$, where $\Delta\rho$ is the density
difference between the fluid and the particles. More precisely, it
has been shown theoretically \cite{Beris1985} and experimentally
\cite{Tabuteau2007} that a single sphere in an infinite yield
stress fluid does not settle as long as
\bea\frac{\tau_y}{\Delta\rho
gd}\geq\frac{1}{21}\label{eq_suspaseuil_stabrepos}\eea

However, it is usually observed that a significant sedimentation
can occur in such materials when they are handled \cite{Cooke2002}. Thus the
question we address here is whether the above stability criterion
still holds for particles in a flowing paste. Whereas stability
and sedimentation at rest have been thoroughly studied in the
literature \cite{Chhabra2006}, these issues have been poorly
addressed in flowing yield stress fluids. The sedimentation of a
single particle in non-Newtonian fluids sheared in a Couette
geometry has been studied by \citet{Gheissary1996}. For all
materials, they attempt to model the sedimentation velocity of the
particle, as a function of the applied shear rate $\gdot$, by the
Stokes velocity $V=(1/18)\Delta\rho gd^2/\eta_{_0}$ in a viscous
fluid of viscosity $\eta_{_0}$ equal to the apparent viscosity
$\eta(\gdot)$ of the sheared fluid. For three different
shear-thinning fluids, they then find that the sedimentation
velocity is significantly lower than expected. This suggests that
the apparent viscosity in the direction of sedimentation is higher
than the apparent viscosity experienced in the Couette flow, a
phenomenon they attribute to the anisotropic character of the
studied materials. The case of a single particle in a Carbopol gel
is also reported; it is worth noting that the particle is not
stable at rest. In this case only, good agreement is found between
the measured and predicted sedimentation velocity.

To our knowledge, sedimentation during a yield stress fluid flow
of particles stable at rest has only been previously studied by
\citet{Merkak2009}, in pipe flows. \citet{Merkak2009} have
observed that particles settle in the sheared fluid in some cases
only; they have rationalized their observations by introducing a
new criterion: they claim that particles do not settle in the
sheared yield stress fluid if \bea\frac{\tau_y}{\Delta\rho g
d}\gtrsim3\label{eq_suspaseuil_stabmerkak}\eea In the cases where
sedimentation is observed, no characterization of the observed
sedimentation velocity is reported.

\begin{figure}[!htb]\begin{center}
\includegraphics[width=10cm]{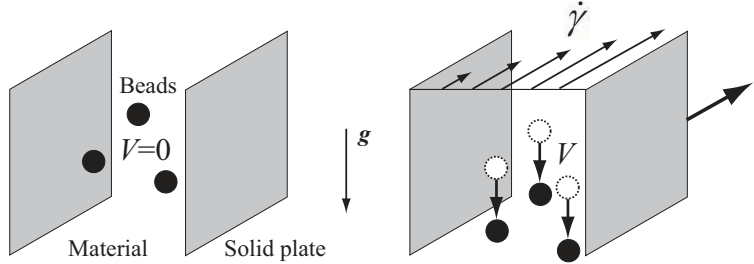}
\caption{Sketch of the experiment.}\label{fig_sketch}
\end{center}\end{figure}

In this paper, we address the question of the impact of a flow on
the possible settling of coarse particles suspended in a yield
stress fluid. Model density-mismatched suspensions of monodisperse
particles are designed to be stable at rest. We focus on a
well-controlled situation: the material is sheared in a coaxial
cylinders (Couette) geometry, which imposes shear in the plane
perpendicular to gravity; the sedimentation flow is thus decoupled
from the shear flow (Fig.~\ref{fig_sketch}). We use Magnetic
Resonance Imaging (MRI) as a noninvasive technique to study the
time evolution of the particle volume fraction during shear. We
first study the situation of low solid fractions for which we
expect that particle interactions are small. Then we study the
case of higher solid fractions, for which collective effects are
known to play a significant role for particles settling in a
simple viscous fluid \cite{Hanratty1957,Davis1985}. We study the
sedimentation velocity as a function of the material properties,
of the particle volume fraction, and of the characteristics of
shear. We then propose a model to account for the observed
features.

\section{Materials and methods}\label{section_materials}

\subsection{Materials}

Most model suspensions we study are suspensions of monodisperse
glass beads in concentrated emulsions; the particle volume
fraction $\phi$ is varied between 5\% and 40\%. We use spherical
glass particles of four different diameters (145, 275, 375, and
405 $\mu$m $\pm7\%$). The emulsions are prepared by dispersing a
100 g/l water solution of CaCl$_2$ in a solution of Span 80
emulsifier (7\%) in dodecane oil at 6000 rpm with a Silverson L4RT
mixer. The average droplet size is 1 $\mu$m; the emulsions are
then viewed by the particles as continuous media
\cite{Mahaut2008a}. In order to study materials of various
rheological properties, we have varied the droplet concentration
between 72 and 85\%.  We have also studied a 5\% suspension of 405
$\mu$m glass beads in a Carbopol gel. The gel is obtained by
dispersing Carbopol 980 (from Noveon) at a 0.3\% concentration in
water at 1000 rpm during 30 min; it is then neutralized with NaOH
at pH=7 and stirred for an entire day to ensure homogeneity of the
material.

\begin{table}[p]\begin{center}\begin{tabular}{|c|c|c|c|c|c|} \hline $\tau_y$ (Pa)&
$\eta_{_{H\!B}}$ (Pa.s$^n$) & $n$ & $d$ ($\mu$m) & $\phi$ (\%) & $Y=\frac{\tau_y}{\Delta\rho gd}$ \\
\hline \hline
\multicolumn{6}{|c|}{\textit{concentrated emulsions}}\\
\hline
8.5 & 3.6 & 0.44 & 275 & 5 & 2.1\\
16.5 & 3.8 & 0.5 & 375 & 5 to 40 & 2.9\\
15.1 & 3.25 & 0.5 & 275 & 5; 10 & 3.7\\
21.4 & 4.25 & 0.5 & 405 & 5 & 3.6\\
25 & 10 & 0.4 & 145 & 5 & 12.1\\
33.2 & 5.0 & 0.5 & 275 & 5 & 8.2\\
33.2 & 5.0 & 0.5 & 405 & 5 & 5.6\\
\hline\hline
\multicolumn{6}{|c|}{\textit{Carbopol gel}}\\
\hline
27.5 & 10.1 & 0.35 & 405 & 5 & 4.7\\
\hline
\end{tabular}\caption{Properties of the studied materials:
yield stress $\tau_y$, consistency $\eta_{_{H\!B}}$ and index $n$
of the yield stress fluids (all material flow curves were fitted
to a Herschel-Bulkley law $\tau=\tau_y+\eta_{_{H\!B}}\gdot^n$);
glass bead diameter $d$; particle volume fraction $\phi$. The
yield number $Y=\frac{\tau_y}{\Delta\rho gd}$ used in stability
criteria (Eqs.~\ref{eq_suspaseuil_stabrepos}
and~\ref{eq_suspaseuil_stabmerkak}) is also
provided.}\label{tab_materials}\end{center}\end{table}

We finally obtain 8 different materials, the properties of which
are displayed in Tab.~\ref{tab_materials}. Due to the specificity
of our setup, which implies to use 1 liter of material per
experiment and does not allow performing many experiments during a
same period, we could not ensure the reproducibility of the
emulsion properties, which explains the variety of the materials.
We were thus able to vary the particle diameter in exactly the
same material only once.

The density of the glass particles is 2.5 whereas the density of
all the studied yield stress fluids is of order 1. Gravity thus
tends to induce sedimentation of the particles. However, all
systems are designed to be stable at rest: their yield number
$Y=\frac{\tau_y}{\Delta\rho gd}$ complies with
Eq.~\ref{eq_suspaseuil_stabrepos} (see Tab.~\ref{tab_materials}),
which ensures that elastic forces exerted by the yield stress
fluid at rest are able to counterbalance the net gravity force.

\begin{figure}[p]\begin{center}
\includegraphics[width=8cm]{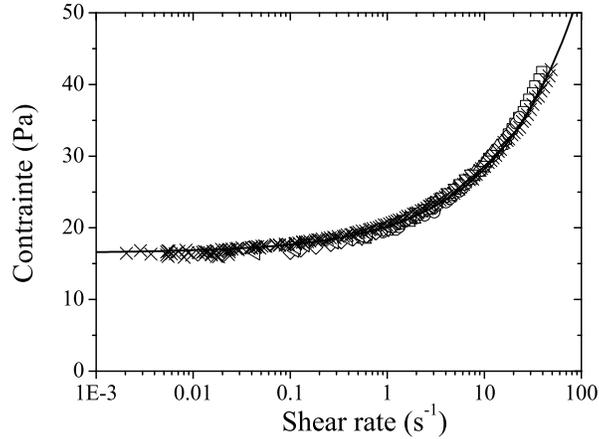}
\caption{Constitutive law $\tau(\gdot)$ of a concentrated emulsion
obtained (i) from macroscopic measurements in a cone-and-plate
geometry (crosses) and (ii) from local MRI measurements in a
Couette geometry (empty symbols; different symbols correspond to
different inner cylinder rotational velocity). The solid line is a
Herschel-Bulkley fit to the macroscopic data
$\tau=\tau_y+\eta_{_{H\!B}}\gdot^{n}$ with $\tau_y=16.5$~Pa,
$\eta_{_{H\!B}}$=3.8~Pa\,s$^{0.5}$, and
$n=0.5$.}\label{fig_rheogram}
\end{center}\end{figure}

\subsection{Rheometry}

The yield stress fluid rheological properties are characterized in
a sandblasted 2\degre cone-and-plate geometry of 40~mm diameter,
with a Bohlin C-VOR 200 rheometer. The material is presheared at a
50~s\m1 shear rate during 60~s. The shear rate $\gdot$ is then
ramped-down from 50~s\m1 to 0.001~s\m1 (logarithmic ramp,
30~s/decade of shear rate), and shear stress vs. shear rate data
$\tau(\gdot$) are recorded during the ramp. We checked that the
materials we study are simple yield stress fluids which do not
display thixotropy, in agreement with previous investigations
\cite{Ovarlez2008,Coussot2009}. $\tau(\gdot)$ data obtained in one
of the concentrated emulsions are displayed in
Fig.~\ref{fig_rheogram}. All material behaviors are well fitted to
a Herschel-Bulkley behavior
$\tau(\gdot)=\tau_y+\eta_{_{H\!B}}\gdot^{n}$ (an example is shown
in Fig.~\ref{fig_rheogram}); the values of the rheological
parameters ($\tau_y$, $\eta_{_{H\!B}}$, $n$) measured on all the
studied materials are displayed in Tab.~\ref{tab_materials}.

\subsection{Sedimentation experiments}

The suspensions are loaded in a Couette geometry, the dimensions
of which are: inner cylinder radius, $R_i=4.1$~cm; outer cylinder
radius, $R_o=6$~cm; height of sheared fluid, $H=11$~cm. Both
cylinders are covered with sandpaper to avoid wall slip. Shear is
induced by the rotation of the inner cylinder at controlled
rotational velocity $\Omega$. The Couette rheometer is inserted in
a Magnetic Resonance Imaging (MRI) setup described in
\cite{Raynaud2002}.

\begin{figure}[!htb]\begin{center}
\includegraphics[width=13.5cm]{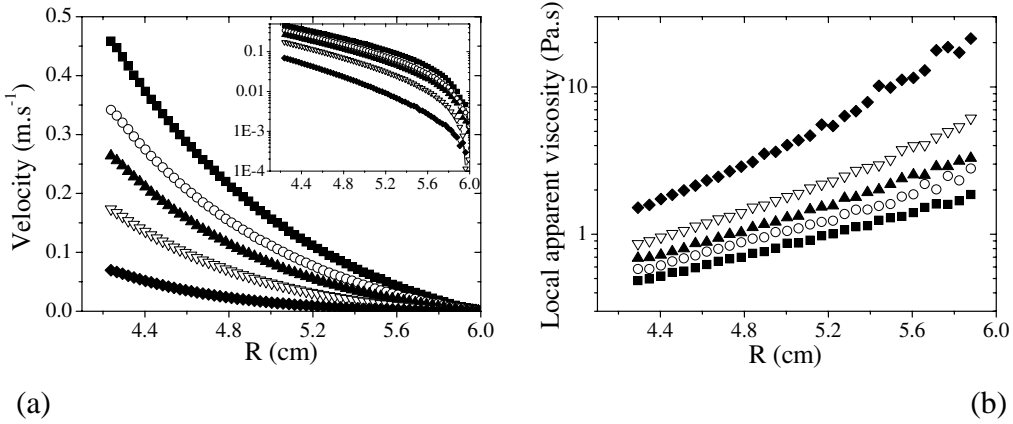}\\
~~~(a)\hfill(b)~~~\\
\caption{(a) Local velocity $v_\theta(r)$ in a 5\% suspension of
275~$\mu$m glass beads in a concentrated emulsion of yield stress
$\tau_y=8.5$~Pa sheared in the gap of a Couette geometry, for
various rotational velocities $\Omega$ of the inner cylinder
(squares: 130~rpm, empty circles: 100 rpm, up triangles: 75 rpm,
empty down triangles: 50 rpm, diamonds: 20 rpm). The inset is a
semi-log plot of the same data. (b) Local apparent viscosity
$\eta[\gdot(r)]$ deduced from the velocity profiles of
Fig.\ref{fig_vitesse_visco_locale}a.}\label{fig_vitesse_visco_locale}
\end{center}\end{figure}

Azimuthal velocity profiles $v_\theta(r)$ are measured using MRI
techniques \cite{Raynaud2002,Rodts2004}; an example is shown in
Fig.~\ref{fig_vitesse_visco_locale}a. The local shear rate
$\gdot(r)$ at a radius $r$ in the gap can then be deduced from
$v_\theta\!(r)$ as $\gdot(r)=v_\theta\!(r)/r-\partial_r
v_\theta\!(r)$; the derivative $\partial_x f$ with respect to
coordinate $x$ of experimental data $f(x_i)$ measured at regularly
spaced positions $x_i$ was here computed as: $\partial_x
f(x_i)=[f(x_{i+1})-f(x_{i-1})]/[x_{i+1}-x_{i-1}]$.

When the material is homogeneous along the vertical direction, the
local shear stress $\tau(r)$ within the gap is obtained from
torque $T$ measurements as $\tau(r)=T/(2\pi r^2H)$. Local data
$\bigl(\tau(r,\Omega),\gdot(r,\Omega)\bigr)$ measured at various
$r$ and various $\Omega$ can finally be combined to obtain the
constitutive law $\tau(\gdot)$ consistent with the observed flows
(more details about this reconstruction technique can be found in
\cite{Ovarlez2008}). Local $\tau(\gdot)$ data measured with this
method in a (pure) concentrated emulsion are displayed in
Fig.~\ref{fig_rheogram}. We observe that there is good agreement
between these local measurements and the macroscopic measurements
obtained in the cone-and-plate geometry. This shows the ability of
the macroscopic measurements to account for the steady-state flow
behavior of the material in our experiments in a wide-gap Couette
geometry.

To investigate shear-induced sedimentation, we study the time
evolution of the $r$-averaged vertical profile $\phi(z)$ of the
particle volume fraction in the suspensions for rotational
velocity $\Omega$ of the inner cylinder ranging between 5 and
100~rpm. The particle volume fraction is obtained both in radial
and vertical directions from density imaging with an accuracy of
0.2\% \cite{Ovarlez2006}. The use of a wide gap allows us to study
large particles. However, it results in unavoidable stress
heterogeneity, as the shear stress distribution is
$\tau(r)=\tau(R_i)R_i^2/r^2$. We thus have to be careful about
shear localization in the experiments \cite{Ovarlez2009}; unless
otherwise noted, all $\Omega$ were high enough to ensure that the
whole gap is sheared (see e.g.
Fig.~\ref{fig_vitesse_visco_locale}a inset). Nevertheless, due to
the nonlinear behavior of the studied materials, the shear rate
and apparent viscosity distributions are strongly heterogeneous.
To illustrate this point, we have computed the local apparent
viscosity $\eta[\gdot(r)]$ of the material in the gap of the
geometry from the interstitial yield stress fluid behavior as
$\eta[\gdot(r)]=\bigl(\tau_y+\eta_{_{H\!B}}\gdot(r)^n\bigr)/\gdot(r)$
and have plotted these values in
Fig.~\ref{fig_vitesse_visco_locale}b in the case of the emulsion
of Fig.~\ref{fig_vitesse_visco_locale}a. It is observed that,
depending on the material and on the boundary conditions, the
shear rate varies by a typical factor of order 10 from the inner
to the outer cylinder, leading to spatial evolution of the
apparent viscosity by a factor of 5.

This heterogeneity is negligible at the particle scale but
important at the suspension scale, and it may have an impact on
the spatial characteristics of sedimentation. This issue is
discussed in Sec.~\ref{subsection_susp_5_profile} and in
the~\ref{appendix_hetero}. To minimize the possible impact of this
heterogeneity, we have measured the vertical concentration
profiles $\phi(z)$ only in a 9~mm thick zone in the gap (from
$r=4.4$~cm to $r=5.3$~cm); in this region, the apparent viscosity
of the material is observed to vary typically by a factor of 2. In
the following, the results are presented as a function of the
spatial average $\bar\gdot$ of $\gdot(r)$ in the measurement
window. In the experiments, $\bar\gdot$ was varied between 3 and
25~s\m1.

All suspensions are stable at rest: the material yield stress is
more than 40 times higher than that fixed by the stability
criterion Eq.~\ref{eq_suspaseuil_stabrepos} (see
Tab.~\ref{tab_materials}). We have checked that the volume
fraction profiles remain indeed homogeneous at rest during 24~h
(Fig.~\ref{fig_suspaseuil_sedimentation}). Moreover, for 5 of the
materials, the yield stress of the interstitial fluid is higher (up to 4
times) than that fixed by the stability criterion under shear
Eq.~\ref{eq_suspaseuil_stabmerkak} proposed by \citet{Merkak2009},
whereas it is lower for 2 of the materials.

In the following, in order to minimize the role of the particle
interactions, we first focus on the behavior of semi-dilute
suspensions of particle volume fraction $\phi=5\%$
(Sec.~\ref{section_susp_5}). We then study the impact of a change
in $\phi$ (Sec.~\ref{section_susp_phi}).

\section{Shear-induced sedimentation in a semi-dilute suspension}\label{section_susp_5}

\subsection{Volume fraction
profiles}\label{subsection_susp_5_profile}

\begin{figure}[p]\begin{center}
\includegraphics[width=5.9cm]{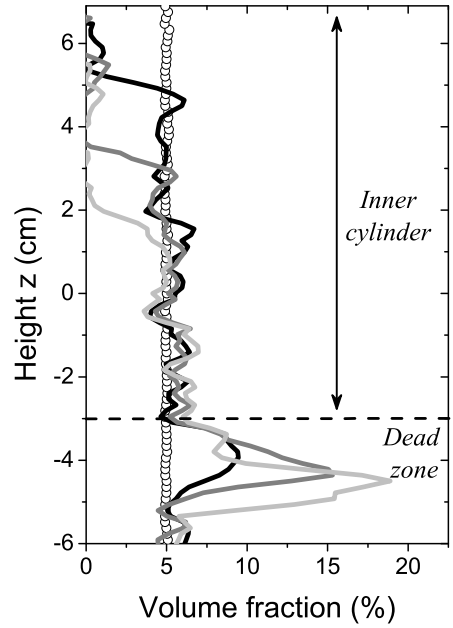}
\caption{Vertical volume fraction profiles observed in the gap of
a Couette geometry in a 5\% suspension of 275~$\mu$m glass beads
in a concentrated emulsion of yield stress $\tau_y=8.5$~Pa, after
a 24~h rest (empty circles) and after 15~min (black line), 30~min
(dark grey line) and 45~min (light grey line) of shear at
$\bar\gdot=4$~s\m1.}\label{fig_suspaseuil_sedimentation}
\end{center}\end{figure}

In Fig.~\ref{fig_suspaseuil_sedimentation}, we show vertical volume
fraction profiles observed at rest and during shear in one of the
studied suspensions of particle volume fraction $\phi=5\%$. The same features are observed in all the systems we have studied.

At rest, the particles appear to remain indefinitely in their
initial position: there is no observable difference between the
vertical concentration profiles measured after loading and after a
24~h rest. Whereas the particles are stable at rest, we observe
that there is sedimentation when the material is sheared. The
sedimentation profiles measured during shear show classical
features of sedimentation in Newtonian fluids \cite{Davies1987}:
the upper part is at a 0\% concentration, the middle part remains
at the initial 5\% concentration, and the particles tend to
accumulate at the bottom of the cup, in the dead zone below the
inner cylinder end. The transition zone between the 0\% and the
5\% regions is rather narrow (with a typical thickness of 6mm),
and defines a sedimentation front that moves continuously towards
the bottom as the flow duration increases.

We do not observe any significant broadening of the sedimentation front in time.
This implies that, although the shear rate distribution is
heterogeneous in the gap (we recall that $\gdot(r)$ varies
typically by a factor of 4 in the measurement zone), there is no
significant spatial heterogeneity of the sedimentation velocity,
which suggests that collective effects are at play. This issue is
discussed in more detail in the~\ref{appendix_hetero}. In the
following, we will thus assume that it is sufficient to consider
the average shear rate $\bar\gdot$ to describe shear-induced
sedimentation.

\begin{figure}[p]\begin{center}
\includegraphics[width=7cm]{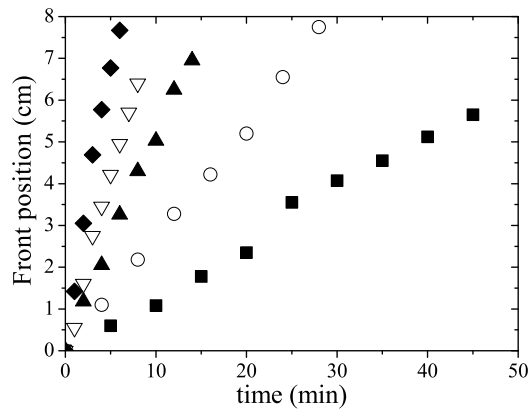}
\caption{Position of the sedimentation front as a function of the
time of shear, for a 5\% suspension of 275~$\mu$m glass beads in
a concentrated emulsion of 8.5~Pa yield stress, for various shear rates: 4~s\m1 (squares), 8.8~s\m1 (empty circles), 14~s\m1 (up triangles),
18.6~s\m1 (empty down triangles) and 25~s\m1
(diamonds).}\label{fig_vit_front}
\end{center}\end{figure}

We observe that the sedimentation front moves linearly in time
(Fig.~\ref{fig_vit_front}). We thus extract the sedimentation
velocity $V$ of the suspension from a linear fit to the front
position vs. time data.

\subsection{Conditions of stability}\label{subsection_susp_5_stability}

The stability at rest of the systems we have studied is consistent with
Eq.~\ref{eq_suspaseuil_stabrepos} (see the values of their yield
number $Y$ in Tab.~\ref{tab_materials}). However, our observations
contrast with previous observations \cite{Merkak2009}: we find
that shear induces sedimentation in all of our systems, at any
imposed shear rate. It thus seems that the proposed stability
criterion Eq.~\ref{eq_suspaseuil_stabmerkak} is not correct (most
of our systems verify Eq.~\ref{eq_suspaseuil_stabmerkak}, see
Tab.~\ref{tab_materials}); from our observations, the possibility
of stability under shear is actually doubtful. The absence of
observable sedimentation in some of the \citet{Merkak2009}
experiments is likely due to the fact that, as shown below in Secs.~\ref{subsection_susp_5_velocity} and~\ref{subsection_susp_5_theory},
the sedimentation velocity can be very small at high material
yield stress and small particle diameter.

\begin{figure}[!htb]\begin{center}
\includegraphics[width=8.5cm]{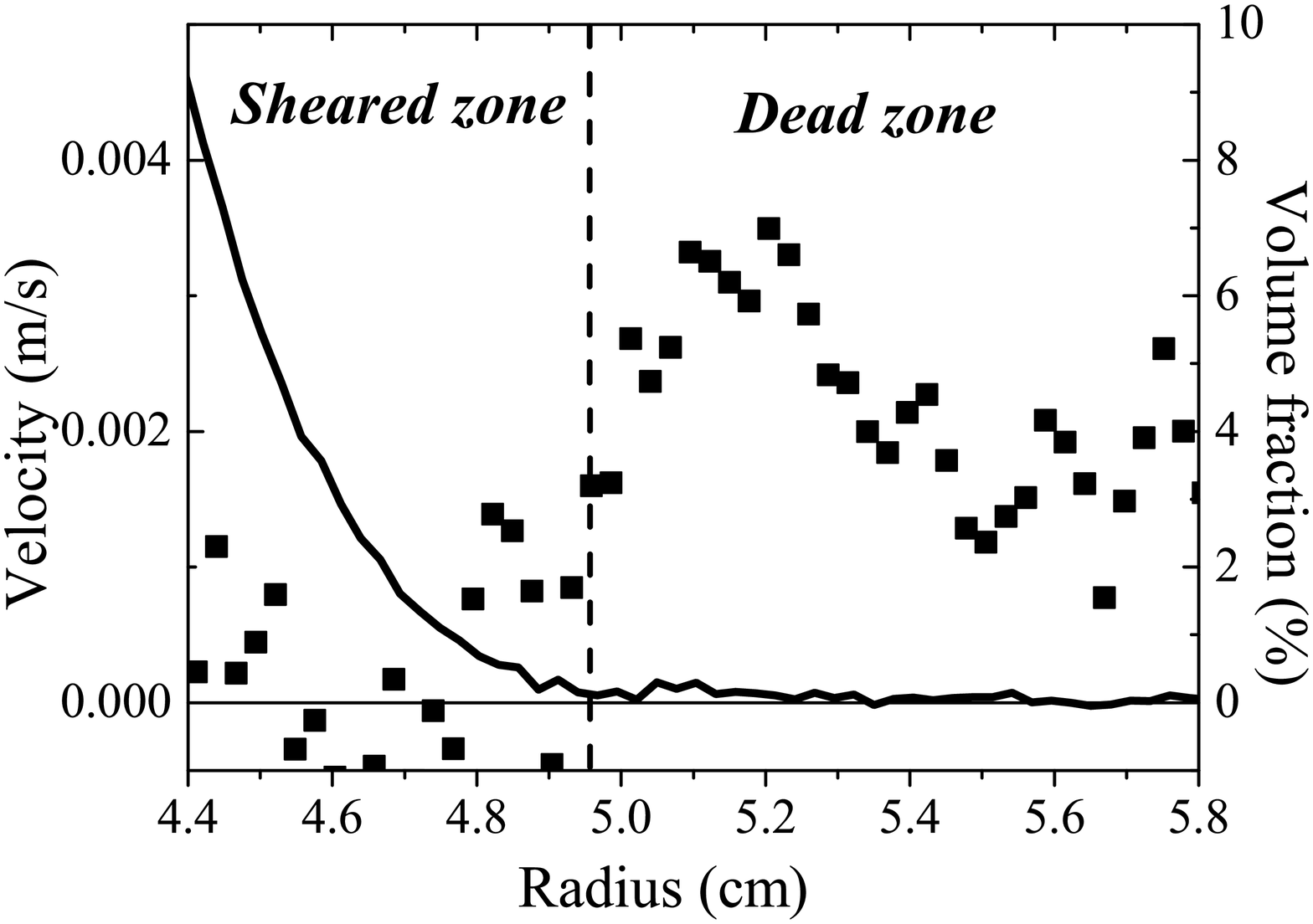}
\caption{Velocity profile (line, left axis) and radial volume
fraction profile (squares, right axis) observed in the gap of a
Couette geometry in a 5\% suspension of 405~$\mu$m glass beads in
a concentrated emulsion of yield stress $\tau_y=21.4$~Pa, after
24~ h of shear at $\Omega=4$~rpm.}\label{fig_localisation}
\end{center}\end{figure}

To illustrate further the role of shear on sedimentation, we have
performed an experiment in which flow is localized
(Fig.~\ref{fig_localisation}), by applying a low rotational
velocity $\Omega=4$~rpm to the inner cylinder on one of the
systems. In these conditions, the material is sheared from the
inner cylinder to the middle of the gap (at $R_c\simeq$ 4.95~cm),
and at rest from $R_c$ to the outer cylinder; the average shear
rate in the sheared region is of the order of 1~s\m1. We recall
that shear localization is here due to the stress heterogeneity,
as the local shear stress value is $\tau(r)=\tau(R_i)R_i^2/r^2$,
which is equal to the yield stress of the material $\tau_y$ at the
interface $R_c$ between the sheared and the unsheared material. We
have sheared the material during 24~h to ensure that all
sedimentation has time to occur if it has to. We have measured the
radial volume fraction profile $\phi(r)$ in the gap of the
geometry after this 24~h shear (Fig.~\ref{fig_localisation}). It
is observed that there are no more particles\footnote{Note that
the measurement fluctuations -- of order 1\% -- are here much
larger for radial profiles than for vertical profiles.} in the
sheared zone, and that the volume fraction in the unsheared zone
is unchanged, consistently with the stability of the system at
rest (see Fig.\ref{fig_suspaseuil_sedimentation}). The interface
between the zones at $\phi=0$\% and $\phi=5$\% exactly corresponds
to the interface between the sheared and the dead zone. This
clearly shows that sedimentation is induced as soon as the
material is sheared, even when the applied stress is very close to
the yield stress (i.e., near $R_c$, where the local shear rate is
close to zero).

\subsection{Sedimentation velocity}\label{subsection_susp_5_velocity}

\begin{figure}[!htb]\begin{center}
\includegraphics[width=13.5cm]{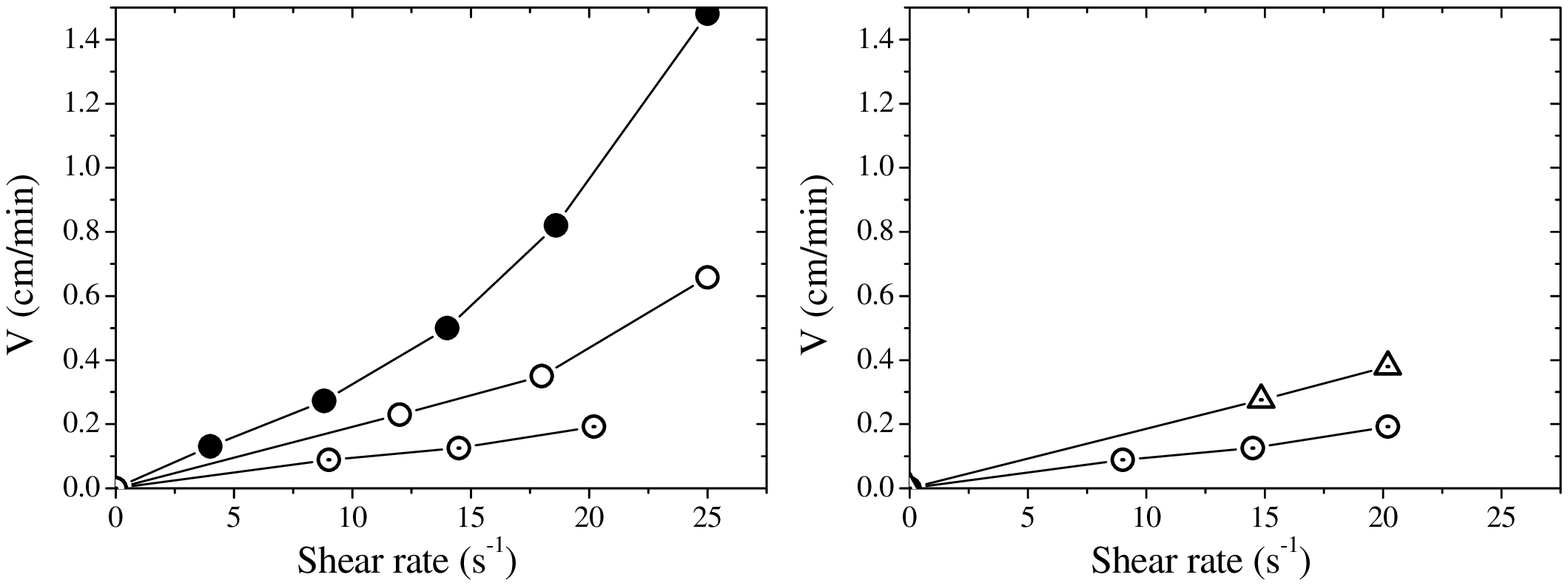}\\
~~~(a)\hfill(b)~~~\\
\caption{(a) Sedimentation velocity $V$ of 275~$\mu$m glass beads
vs. shear rate $\bar\gdot$ in three different concentrated emulsions, of
yield stress 8.5 Pa (filled circles), 15 Pa (empty circles) and 33
Pa (dotted circles). (b) Sedimentation velocity $V$ of 275~$\mu$m
(dotted circles) and 405~$\mu$m (dotted triangles) glass beads vs.
shear rate $\bar\gdot$ in a concentrated emulsion of 33 Pa yield stress. The
particle volume fraction of the suspensions is
$\phi=5\%$.}\label{fig_vit_sedim_d_et_tau}
\end{center}\end{figure}

In order to better understand shear-induced sedimentation, we now
compare the sedimentation velocities $V$ observed in all systems under various
conditions. Our observations can be summarized as follow:
\begin{itemize}
\item $V$ increases with the applied shear rate $\bar\gdot$
(Fig.~\ref{fig_vit_sedim_d_et_tau}) \item for a given particle
diameter $d$, at a given $\bar\gdot$, $V$ is a decreasing function
of the material yield stress $\tau_y$
(Fig.~\ref{fig_vit_sedim_d_et_tau}a). \item in a yield stress
fluid of given rheological properties, $V$ is an increasing
function of the particle diameter $d$
(Fig.~\ref{fig_vit_sedim_d_et_tau}b)
\end{itemize}
See also Fig.~\ref{fig_vit_sedim_suspaseuil} where data obtained
on all the studied materials are shown.\\

These results can be qualitatively understood: $V$ is actually
expected to decrease when the plastic to gravity stress ratio
$Y=\frac{\tau_y}{\Delta\rho gd}$ increases. Moreover, the increase
of $V$ with $\bar\gdot$ suggests that the settling particles are
sensitive to the apparent viscosity
$\eta(\bar\gdot)=\tau(\bar\gdot)/\bar\gdot$ of the sheared
material, which decreases with $\bar\gdot$ for all the studied
materials.

\subsection{Theoretical analysis}\label{subsection_susp_5_theory}
In this section, we aim to better understand the phenomenon of
shear-induced sedimentation, and to predict the value of the
sedimentation velocity $V$ as a function of the material and flow
characteristics. We point out that the particle Reynolds number $Re_p$ for both the shear flow ($\rho d^2 \gdot/\eta(\gdot)$) and the settling flow ($\rho d V/\eta(\gdot)$) is of order $10^{-3}$ or less in all experiments; our analysis is thus made in the Stokes regime. In the following, we first deal with the problem of
the sedimentation of a single sphere of diameter $d$ in the
sheared yield stress fluid; we will comment on the volume fraction
dependence of $V$ below.

Shear-induced sedimentation involves a complex 3D flow of the
yield stress fluid around the particle. We thus have to consider a
tensorial form of the Herschel-Bulkley behavior; here we assume that the material behavior obeys
its isotropic form \cite{Oldroyd1950,Coussot2005,Ovarlez2010}
\bea\tij=2[(\tau_y+\eta_{_{H\!B}}\gdot^n)/\gdot]\dij\label{eq_HB}\eea
where $\tij$ is the deviatoric stress tensor, $\dij$ is the strain rate tensor (i.e. the symmetric part of the velocity gradient), and the shear rate is $\gdot=\sqrt{2\dij\dij}$. This form has
been shown to be consistent with recent observations of complex
yield stress fluid flows \cite{Ovarlez2010,Rabideau2010}.

When two orthogonal flows\footnote{We recall that two flows of
strain rate tensors $d^1_{ij}$ and $d^2_{ij}$ are said to be orthogonal
when $d^1_{ij}\,d^2_{ij}=0$.} are superimposed, it has been shown
that, if the strain rate of one of the flows is much higher than
that of the other flow, the flow resistance to the secondary flow
is purely viscous (linear) and is characterized by an effective
viscosity equal to the apparent viscosity $\eta(\gdot)$ of the
main flow \cite{Ovarlez2010}. This can be understood for the
isotropic Herschel-Bulkley law (Eq.~\ref{eq_HB}): e.g, if we
consider a main simple shear flow of strain rate $\drt$ in a
Couette geometry, the shear resistance $\trz$ to a secondary
orthogonal flow in the $z$ direction, of strain rate $\drz$ such
that $\drz~\ll~\drt$, is \bea
\trz\simeq2[(\tau_y+\eta_{_{H\!B}}(2\drt)^n)/2\drt]\,\drz=2\eta(\gdot)\,\drz\label{eq_HB_rz}\eea
with $\gdot\simeq2\drt$: the flow resistance to this secondary
flow in the vertical direction is thus characterized by a viscous
behavior of effective viscosity
$\eta(\gdot)=[(\tau_y+\eta_{_{H\!B}}\gdot^n)/\gdot]$ fixed by the
characteristics of the Couette flow only.

In all of our experiments, the settling flow can be considered as
a secondary flow as compared to the shear flow. Indeed, its
characteristic shear rate $\gdot_s$ is of order $V/d$; this value
is here 50 to 400 times lower than $\gdot\simeq2\drt$. In these conditions,
given the above remarks about the flow resistance to secondary
flows, it is tempting to model the sedimentation velocity of a
single sphere in the sheared yield stress fluid by the velocity
$V_\text{Stokes}$ of a sphere in a Newtonian medium of viscosity
given by the apparent viscous resistance
$\eta(\gdot)=\tau(\gdot)/\gdot$ to the shear flow (here,
$\tau=\trt$ and $\gdot=2\drt$ characterize the shear flow induced
by the rotation of the inner cylinder of the Couette cell):\bea
V_\text{Stokes}=\alpha_S\frac{\Delta\rho g
d^2}{\eta(\gdot)}\label{eq_vit_stokes}\eea with $\alpha_S=1/18$.

However, this preliminary analysis does not account for the
complexity of flow around a sphere: it is based on
\textit{orthogonal} superimposed flows, which is not the case of
the shear and sedimentation flows around a sphere. In this last
case, one cannot simply describe the shear resistance to
sedimentation as in Eq.~\ref{eq_HB_rz}, because the strain rate tensor field around the sphere is not the sum of two orthogonal tensor fields
(i.e., $\dG:\ds\neq0$ in Eq.~\ref{eq:taus}). Moreover, the contribution to the drag
force of the normal stresses exerted on the particle also have to
be properly taken into account in the analysis. In
the~\ref{appendix_theory}, a rigorous theoretical analysis of the
problem of shear-induced sedimentation is made, in the case where
the sedimentation flow is a secondary flow (i.e., when $V/d \ll
\gdot$ ). It is shown that the scaling proposed in
Eq.~\ref{eq_vit_stokes} remains valid in two limit cases, for low
and high inverse Bingham number
$Bi$\m1$=\eta_{_{H\!B}}\gdot^n/\tau_y$.\\ When $Bi$\m1$\ll1$
(`plastic' flows), it is expected that: \bea
V=\alpha_p\frac{\Delta\rho g
d^2}{\tau_y/\gdot}\label{eq_vit_stokes_plas} \eea When
$Bi$\m1$\gg1$ (`viscous' flows), it is expected that:\bea
V=\alpha_v\frac{\Delta\rho g
d^2}{\eta_{_{H\!B}}\gdot^{n-1}}\label{eq_vit_stokes_visc} \eea
Note that the constants $\alpha_p$ and $\alpha_v$ are unknown and
are \textit{a priori} different, which means that
Eq.~\ref{eq_vit_stokes} cannot be correct and that in a general
case, $V\not\propto 1/\eta(\gdot)$. Nevertheless, as it gives the
correct scaling predictions in both the `plastic' and `viscous'
limits, we will compare our results to Eq.~\ref{eq_vit_stokes} in
the following.

These last equations are expected to be valid for a single sphere
only, a case we cannot study with our means of investigation. We
recall that our first results concern here semi-dilute suspensions
of volume fraction $\phi=5$\%. For a suspension of monodisperse
spheres of volume fraction $\phi$, by analogy with viscous
suspensions, we expect the sedimentation velocity $V(\phi)$ of the
suspension to be equal to that of a single sphere multiplied by a
decreasing function of $\phi$ only, the hindrance function
$f(\phi)$, with $f(0)=1$ \cite{Hanratty1957,Davis1985} (we will
come back onto this point in Sec.~\ref{section_susp_phi}).

\subsection{Comparison experiment/theory}\label{subsection_susp_5_analysis}

\begin{figure}[!htb]\begin{center}
\includegraphics[width=13.5cm]{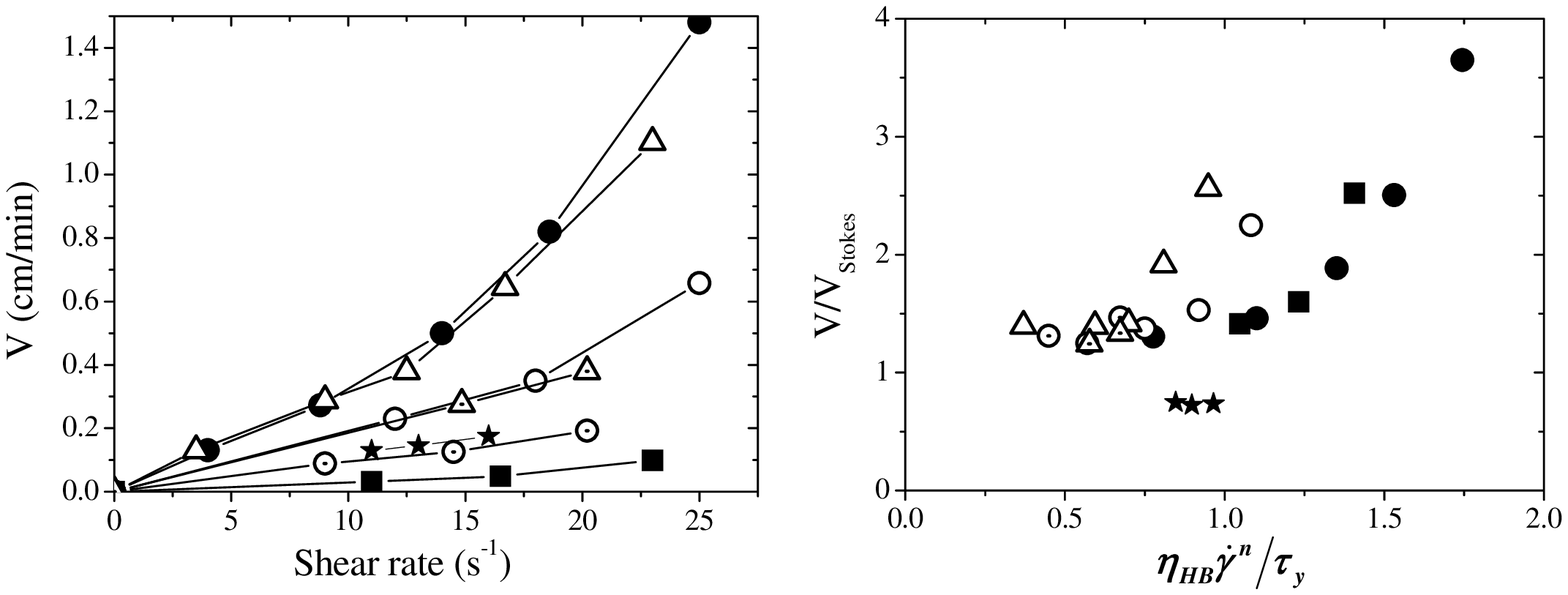}\\
~~~(a)\hfill(b)~~~\\
\caption{(a) Sedimentation velocity $V$ of glass beads suspended
at a 5\% volume fraction in sheared yield stress fluids as a
function of the applied shear rate $\bar\gdot$, for various bead
diameter $d$, various concentrated emulsions (filled squares: d=145~$\mu$m,
$\tau_y=25$~Pa; circles: d=275~$\mu$m, $\tau_y=8.5$~Pa (filled),
15~Pa (empty), 33~Pa (dotted); triangles: d=405~$\mu$m,
$\tau_y=21.5$~Pa (empty), 33~Pa (dotted)) and a Carbopol gel
(stars: d=405~$\mu$m, $\tau_y=27.5$~Pa). (b) Same data rescaled by
the velocity $V_\text{Stokes}=(\Delta \rho g{d^2})/(18\eta)$ of a
single sphere that would fall in a Newtonian medium of viscosity
$\eta$ equal to the apparent viscosity $\eta(\bar\gdot)$ of the
pure sheared yield stress fluid; data are displayed vs. the
inverse Bingham number
$Bi$\m1$=\eta_{_{H\!B}}\bar\gdot^n/\tau_y$.}\label{fig_vit_sedim_suspaseuil}
\end{center}\end{figure}

In order to test the above theoretical analysis, we show the data
obtained in all materials in Fig.~\ref{fig_vit_sedim_suspaseuil}a,
and we plot all the measured sedimentation velocities $V$ rescaled
by the velocity $V_\text{Stokes}$ given by Eq.~\ref{eq_vit_stokes}
in Fig.~\ref{fig_vit_sedim_suspaseuil}b. We recall that
$V/V_\text{Stokes}$ is expected to tend towards two different
constant values at low and large inverse Bingham number
$Bi$\m1$=\eta_{_{H\!B}}\bar\gdot^n/\tau_y$. $V/V_\text{Stokes}$ is
thus plotted here vs. $Bi$\m1.

For shear-induced sedimentation in the concentrated emulsions, we
first observe that all the data fall onto a master curve
$V/V_\text{Stokes}\simeq1.4$ up to $Bi$\m1$\simeq1$, in good
agreement with the theory in the `plastic' regime. It is worth noting that the parameters $d^2$, $\tau_y$ and $\bar\gdot$ involved in the scaling of Eq.~\ref{eq_vit_stokes} were here varied independently by a factor of 8, 4 and 4, respectively.

For higher
values of $Bi$\m1, $V/V_\text{Stokes}$ is then found to increase
with $Bi$\m1; it can reach values that are up to 3 times higher than at low
$Bi$\m1 value. The existence of such a regime could be expected
from the theory as $V/V_\text{Stokes}$ \textit{a priori} tends
towards different values at low and high $Bi$\m1. However, we do
not observe $V/V_\text{Stokes}$ to tend towards a constant value
when increasing $Bi$\m1, in contrast with what is expected in the
`viscous' regime. Nevertheless, it should be noted that the higher
value of $Bi$\m1 tested here was of order 1.8; further experiments
at higher $Bi$\m1 values would be needed to really test the theory
in the `viscous' regime. Data obtained here for $Bi$\m1$\gtrsim1$
do not fall along a master curve, which means that the
Bingham number should not be the only relevant parameter to consider to describe this intermediate regime.

For shear-induced sedimentation in the Carbopol gel, the important
result is that the proposed scaling with the apparent viscosity of
the sheared material $\eta(\gdot)$ seems to remain correct in the
`plastic' regime, as $V/V_\text{Stokes}$ does not depend on
$\bar\gdot$ for $Bi$\m1$<1$. Nevertheless, we now find a different
value $V/V_\text{Stokes}\simeq0.7$. It thus seems that
shear-induced sedimentation is also sensitive to aspects of the
material rheological behavior that we have not considered in the
analysis. A possibility is that normal stress differences play a
role, since they can be large in sheared Carbopol gels
\cite{Piau2007}, and since they are not taken into account in the
expression of the Herschel-Bulkley law that we use.

For the sake of comparison with the properties of sedimentation in
Newtonian fluids, we have performed sedimentation experiments with
a 5\% suspension of 275 $\mu$m glass beads in a Newtonian oil
(viscosity 2.6 Pa.s), in several situations: (i) in the gap of the
Couette cell at rest, (ii) in the gap of the Couette cell when the
suspension is sheared, and (iii) in a large cylindrical container
of 12~cm diameter and of 10~cm height. In the gap of the Couette
cell, we have observed $V/V_\text{Stokes}\simeq1.2$, both at rest
and when the fluid is sheared (as expected from the linearity of
the fluid behavior). In the cylindrical container, we have
observed that $V/V_\text{Stokes}\simeq0.7$ (note that
$V/V_\text{Stokes}\simeq0.8$ is expected at $\phi=5\%$ in a
Newtonian fluid in an `infinite' geometry \cite{Hanratty1957}).
This shows that the characteristics of sedimentation in a
Newtonian fluid are affected by the characteristics of the
geometry we use\footnote{This dependence could come, e.g., from
the curvature of the geometry \cite{Beenaker1985}, from a small
tilt angle \cite{Davis1985}, or from the limited size of the gap;
we note in particular that the gap size is here of the order of
the correlation length $\simeq10 d \phi^{-1/3}$ of the particle
settling velocities \cite{Guazzelli2011}, and that these
fluctuations play a role in the backflow contribution to the
suspension sedimentation velocity
\cite{Guazzelli2011,Brenner1999}.}; it is also most probably the
case for shear-induced sedimentation in a yield stress fluid. The
value of $V/V_\text{Stokes}$ we measure here would thus be
difficult to compare to the results of analytic or numerical
computations in plastic materials under homogeneous simple shear.

Note also that in Eq.~\ref{eq_vit_stokes} the sedimentation
velocity is supposed to be set by the local value $\eta[\gdot(r)]$ of the apparent viscosity of the sheared fluid, and thus to depend on the radial position $r$ in the gap following
$V(r)\propto1/\eta[\gdot(r)]$. In Fig.~\ref{fig_vitesse_visco_locale}b we
have shown that there are large $\eta[\gdot(r)]$ variations in the
gap: it varies typically by a factor of 5 in the whole gap and by a factor of 2 in the measurement zone,
which should imply that, when averaging the sedimentation velocities in this zone, the sedimentation front corresponding to
$\phi=5$\% in this region should move 2 times faster than the
front corresponding to $\phi$=0\%. This is not the case
experimentally as we have observed no broadening of the
sedimentation front (Fig.~\ref{fig_suspaseuil_sedimentation}),
which suggests that collective effects are at play
(see~\ref{appendix_hetero} for more details). This raises the
question of the relevant value of $\eta[\gdot(r)]$ to take into
account in the theoretical analysis of the sedimentation velocity.
Here we have no answer to this question: it could be the minimum
or maximum value of $\eta[\gdot(r)]$ in the gap, its average, etc.  We have thus made
an arbitrary choice by choosing $\eta(\bar\gdot)$. We checked that
this choice is not crucial: e.g., putting the maximum or minimum
value of $\eta[\gdot(r)]$ in the gap in the analysis does not
change the features observed experimentally (i.e., the existence
of a plateau of $V/V_\text{Stokes}$ at low $Bi$\m1 and an increase
of $V/V_\text{Stokes}$ for $Bi$\m1$\gtrsim1$). However, this
choice affects the quantitative value of $V/V_\text{Stokes}$,
which, again, would make it impossible to make a quantitative
comparison to the results of analytic or numerical computations in
plastic materials under homogeneous simple shear.\\

To conclude, we have shown that the sedimentation velocity of
spherical particles at low solid fraction in a sheared yield
stress fluid can be fairly predicted with Eq.~\ref{eq_vit_stokes}
in the `plastic' regime
($Bi$\m1$=\eta_{_{H\!B}}\gdot^n/\tau_y<1$). The drag constant
$\alpha$ seems to be close to that observed in Newtonian fluids,
but its exact value cannot be directly derived from our
experimental measurements as it depends on the characteristics of
the geometry we use. The existence of a `viscous' regime where the
scaling of Eq.~\ref{eq_vit_stokes} should remain valid could not
be tested with our systems; nevertheless, our results (the  increase of
$V/V_\text{Stokes}$ with $Bi$\m1 for $Bi$\m1$\gtrsim1$) suggest
that the drag constant $\alpha$ is significantly higher in this
regime than in the `plastic' regime. In the next section, we study
the dependence of the sedimentation velocity on the solid fraction
in the `plastic' regime.

\section{Dependence of the sedimentation velocity on the particle volume fraction}\label{section_susp_phi}

\subsection{Observations}

In a second stage, we have studied the impact of a change in the
particle volume fraction $\phi$ on the sedimentation velocity. We
have studied suspensions of 375 $\mu$m beads in a concentrated emulsion, with
$\phi$ ranging from 5\% to 40\% (see Tab.~\ref{tab_materials}). A
low macroscopic shear rate $\bar\gdot$=8 s\m1 was imposed in order
to ensure that we are in the simplest case (`plastic' regime)
where Eq.~\ref{eq_vit_stokes} is valid.

\begin{figure}[!htb]\begin{center}
\includegraphics[width=7.5cm]{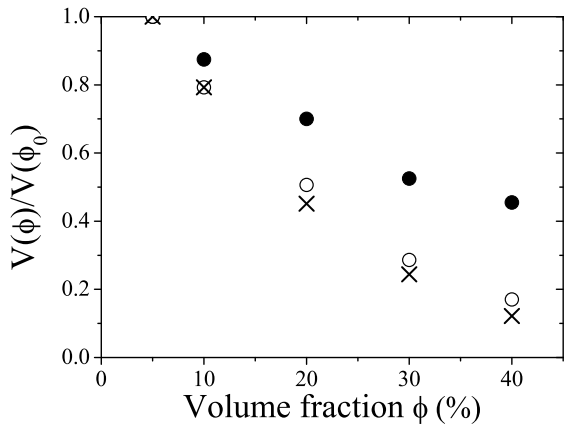}
\caption{Dimensionless sedimentation velocity
$V(\phi)/V(\phi_{_0})$ vs. particle volume fraction $\phi$, where
$\phi_{_0}=5\%$, for suspensions sheared in a Couette geometry at a
$\bar\gdot=8$~s\m1 macroscopic shear rate (filled circles) and for
viscous suspensions (crosses, data from \cite{Hanratty1957}). We
also plot the dimensionless velocity corrected for the impact of
the volume fraction change on the apparent viscosity of the interstitial
fluid (empty circles):
$V(\phi)/V(\phi_{_0})\times\eta[\gdot_l(\phi)]/\eta[\gdot_l(\phi_{_0})]$,
where $\eta[\gdot_l(\phi)]$ is the apparent viscosity of the
sheared interstitial yield stress fluid at a local shear rate
$\gdot_l(\phi)$ computed following
Eq.~\ref{eq_vit_stokes_phib}.}\label{fig_suspaseuil_sedim_phi}
\end{center}\end{figure}

The sedimentation velocities $V(\phi)$ measured at a same
macroscopic shear rate $\bar\gdot$ are displayed in
Fig.~\ref{fig_suspaseuil_sedim_phi}, rescaled by the velocity
$V(\phi_{_0})$ measured for $\phi_{_0}=5\%$. These values are
compared to the dimensionless sedimentation velocity
$V(\phi)/V(\phi_{_0})$ observed in viscous suspensions in the
literature \cite{Hanratty1957}, which accounts for collective
effects in Newtonian materials.

It is observed that the sedimentation velocity decreases very
moderately with the volume fraction in a sheared yield stress
fluid, whereas a strong decrease is observed in Newtonian fluids.
E.g., at a 40\% particle volume fraction, the sedimentation
velocity in the sheared yield stress fluid decreases by only a
factor of 2 as compared to a 5\% suspension, which is 4 times less
important than in a Newtonian fluid.

\subsection{Theoretical analysis}

This observation can be understood by analyzing the behavior of
the material at the local scale. Indeed, as the particles are
rigid, shear is more and more concentrated in the fluid between
neighboring particles as the particles get closer. This implies
that, at a given macroscopic
shear rate $\bar\gdot$, the local shear rate $\gdot_l$ in the
interstitial fluid is an increasing function of the particle volume fraction $\phi$. Then, when $\phi$ is increased, the apparent
viscosity
$\tau(\gdot_l)/\gdot_l=\tau_y/\gdot_l+\eta_{_{H\!B}}/\gdot_l^{1-n}$
of the interstitial fluid decreases, i.e. the viscous
resistance to sedimentation decreases, which may explain why
$V(\phi)/V(\phi_{_0})$ is higher in a sheared yield stress fluid than in a viscous fluid of constant viscosity.

The value of the average local shear rate $\bar\gdot_l$ can be
estimated as a function of the macroscopic shear rate $\bar\gdot$
and of the particle volume fraction $\phi$, by comparing the
macroscopic properties of the suspension (elastic modulus
$G'(\phi)$, yield stress $\tau_y(\phi)$, consistency
$\eta_{_{H\!B}}(\phi)$) and those of the suspending fluid
($G'(0)$, $\tau_y(0)$, $\eta_{_{H\!B}}(0)$) \cite{Chateau2008}.
E.g., the elastic energy stored below the yield stress in the
suspension at the macroscopic scale $G'(\phi)\bar\gdot^2$ must
match the elastic energy stored at the microscopic scale, which is
stored in the interstitial fluid only and is
$(1-\phi)G'(0)\bar\gdot_l^2$. This implies that
$\bar\gdot_l(\phi)=\bar \gdot\sqrt{g(\phi)/(1-\phi)}$ where
$g(\phi)=G'(\phi)/G'(0)$ is the linear response of the material
\cite{Mahaut2008a,Chateau2008,Vu2010}. A classical expression for
$g(\phi)$ is the Krieger-Dougherty law
$g(\phi)=(1-\phi/\phid)^{-2.5\phid}$, which is in agreement with
the experimentally observed behavior \cite{Mahaut2008a} for
monodisperse suspensions. For sheared suspensions of monodisperse
spheres, a reliable value of $\phid=60.5\%$ was recently measured
at a local scale \cite{Ovarlez2006}. This finally means that the
average local shear rate in the interstitial fluid can be
estimated with no fitting parameter as
$\bar\gdot_l(\phi)=\bar\gdot\sqrt{(1-\phi/\phid)^{-2.5\phid}(1-\phi)^{-1}}$
with $\phid=60.5\%$, in agreement with previous experimental data
\cite{Mahaut2008a,Chateau2008,Vu2010}.

We thus propose to model the sedimentation velocity $V(\phi)$ of a
suspension of volume fraction $\phi$ by taking into account the
local viscosity $\eta[\bar\gdot_l(\phi)]$ of the interstitial
fluid into Eq.~\ref{eq_vit_stokes}, and by
multiplying this last equation, valid for a single sphere only, by
the same ``hindrance function'' $f_\text{Newt.}(\phi)$ as in a
Newtonian fluid to account for collective effects
($f_\text{Newt.}(\phi)$ is the sedimentation velocity of a
Newtonian suspension rescaled by the Stokes velocity). This
finally leads to the following set of equations: \bea
V(\phi)=\alpha\frac{\Delta\rho
g d^2}{\eta[\bar\gdot_l(\phi)]} f_\text{Newt.}(\phi)\label{eq_vit_stokes_phi}\\
\text{where}\quad
\bar\gdot_l(\phi)=\bar\gdot\sqrt{(1-\phi/\phid)^{-2.5\phid}(1-\phi)^{-1}}\label{eq_vit_stokes_phib}\eea
with $\phid=60.5\%$, where $\eta(\gdot)=\tau(\gdot)/\gdot$ is the
apparent viscosity of the pure yield stress fluid sheared at a
$\gdot$ shear rate. We recall that Eq.~\ref{eq_vit_stokes_phi} is
\textit{a priori} valid at low and large $Bi$\m1 values only, and
that the drag constant $\alpha$ is expected to take different
values in the `plastic' and `viscous' regimes, and to be close to the
value $1/18$ observed in Newtonian materials in the `plastic' regime
(see Sec.~\ref{subsection_susp_5_analysis}). It is worth noting
that there is no fitting parameter in
Eqs.~\ref{eq_vit_stokes_phi},~\ref{eq_vit_stokes_phib}.

In order to test this model, we report in
Fig.~\ref{fig_suspaseuil_sedim_phi} the values of
$V(\phi)/V(\phi_{_0})\times\eta[\bar\gdot_l(\phi)]/\eta[\bar\gdot_l(\phi_{_0})]$,
where $\phi_{_0}=5\%$, which should be equal to the dimensionless
sedimentation velocity of a Newtonian fluid
$f_\text{Newt.}(\phi)/f_\text{Newt.}(\phi_{_0})$ from
Eq.~\ref{eq_vit_stokes_phi}. A fair agreement is observed between
our theoretical expression and the values
$f_\text{Newt.}(\phi)/f_\text{Newt.}(\phi_{_0})$ measured in
Newtonian suspensions \cite{Hanratty1957}.

\section{Conclusion}

We have addressed the question of the impact of a flow on the
stability of coarse particles suspended in yield stress fluids. In
contrast with previous investigation \cite{Merkak2009}, we
have observed that shear induces sedimentation of the particles in
all the studied systems, which were stable at rest. It thus
seems that any density-mismatched suspension of particles or
bubbles \cite{Goyon2010} in a yield stress fluid will tend to
become heterogeneous when it is transported. It makes it
particularly important to model shear-induced sedimentation as a
function of the characteristics of shear and of the material
properties.

At low shear rate (`plastic regime',
$Bi$\m1$=\eta_{_{H\!B}}\gdot^n/\tau_y<1$), we have proposed a
quantitative prediction of the sedimentation velocity, which is in
good agreement with our experimental observations. We have shown
that shear-induced sedimentation can be modelled by considering
that the interstitial fluid behaves like a viscous fluid -- of
viscosity equal to the apparent viscosity $\tau(\gdot)/\gdot$ of
the sheared fluid -- in the direction orthogonal to shear. The
sedimentation velocity of spherical particles in a sheared yield
stress fluid can then be fairly predicted using
Eq.~\ref{eq_vit_stokes_phi}, with a drag constant $\alpha$ close
to that observed in Newtonian media. An increase of the particle
volume fraction $\phi$ plays here two contradictory roles: it
hinders settling, which is accounted for by the same hindrance
function $f(\phi)$ as in Newtonian fluids, and it decreases the
viscous resistance of the interstitial fluid because of shear
concentration between the particles, which is accounted for by
estimating the local shear rate in the interstitial yield stress
fluid with Eq.~\ref{eq_vit_stokes_phib}
\cite{Mahaut2008a,Chateau2008}.

At this stage, this modelling has been shown to be valid at proximity of
the yield stress only. The existence of a `viscous' regime
($Bi$\m1$>>1$) where the scaling of Eq.~\ref{eq_vit_stokes_phi}
should remain valid is predicted theoretically, but we could not
test it experimentally. Nevertheless, our results suggest that
the drag constant $\alpha$ should be significantly higher (by a factor of
3 or more) in this regime than in the `plastic' regime, which
would result in high shear-induced sedimentation velocities. The
behavior at high shear rates thus remains to be fully
characterized and understood.

\appendix

\section{Impact of the shear rate
heterogeneity}\label{appendix_hetero}

In the following, we study the possible impact of the shear rate
heterogeneity on the spatial characteristics of shear-induced
sedimentation.

Let us assume, consistently with our findings in the `plastic'
regime (Sec.~\ref{subsection_susp_5_analysis}), that particles
situated at a given radial position $r$ in the gap settle in the
sheared yield stress fluid as in a viscous medium of local
viscosity $\eta(r)=\tau(r)/\gdot(r)$, i.e. that the local
sedimentation velocity is $V(r)\propto1/\eta(r)$. It is thus
assumed here that the local sedimentation velocity $V(r)$ of the
suspension at a given position is independent of the sedimentation
velocity in its surroundings. The theoretical volume fraction
profile $\bar\phi(z,t)$ corresponding to the experimental conditions
after a given time $t$ of shear can then be easily computed by
averaging the theoretical volume fraction profile $\phi(r,z,t)$ at
a given vertical position $z$ from $r=R_1=4.4$~cm to $R_2=5.3$~cm.

\begin{figure}[!htb]\begin{center}
\includegraphics[width=6cm]{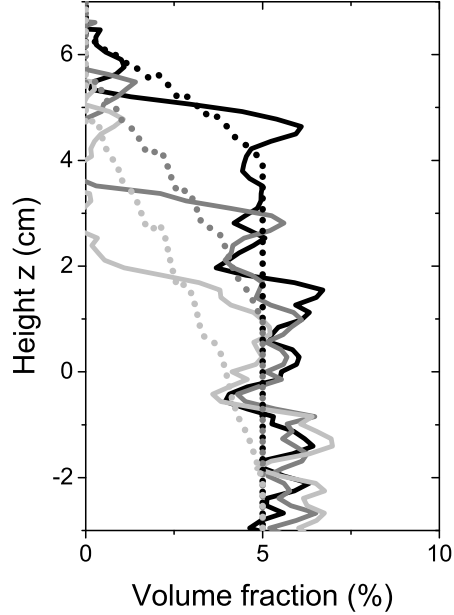}
\caption{Vertical volume fraction profiles observed in the gap of
a Couette geometry in a 5\% suspension of 275~$\mu$m glass beads
in a concentrated emulsion of yield stress $\tau_y=8.5$~Pa after
15~min (black line), 30~min (dark grey line) and 45~min (light
grey line) of shear at $\bar\gdot=4$~s\m1. The dotted lines are
the theoretical profiles expected from Eq.~\ref{eq_vit_stokes} by
taking into account the heterogeneity of the apparent viscosity in
the sheared material, under the assumption that the
sedimentation velocity of the suspension at a given radial
position is set by the local viscosity $\eta(r)=\tau(r)/\gdot(r)$
of the sheared yield stress fluid only, independently of the
sedimentation velocity in its
surroundings.}\label{fig_profils_vs_theorie}
\end{center}\end{figure}

To do that, defining $z=h$ as the top of the suspension, we start
at time $t=0$ with a homogeneous suspension of volume fraction
$\phi_{_0}=5\%$, i.e., we write
$\phi(r,z,t=0)=\phi_{_0}\,\text{H}(h-z)$ with $\text{H}$ the
Heaviside function ($\text{H}(z)=1$ if $z\geq0$ and
$\text{H}(z)=0$ if $z<0$). After a time $t$ of shear at a given
macroscopic shear rate, the local volume fraction is then computed
as $\phi(r,z,t)=\phi_{_0}\,\text{H}(h-V(r)t-z)$, with  the local
sedimentation velocity $V(r)\propto1/\eta(r)$ calculated from the
experimentally measured value of $\eta(r)$ (see
Fig.~\ref{fig_vitesse_visco_locale}b). The profile $\bar\phi(z,t)$
that should theoretically be measured in these conditions is then
the average of $\phi(r,z,t)$ between the radial positions $R_1$
and $R_2$ (the measurement window has a constant extent in the
azimuthal direction) and is
$\bar\phi(z,t)=\frac{1}{R_2-R_1}\int^{R_2}_{R_1}\phi_{_0}\,\text{H}(h-V(r)t-z)\,\text{d}r$.
Such profiles are compared to experimental profiles in
Fig.~\ref{fig_profils_vs_theorie}.

There is strong discrepancy between what is expected from this
first naïve analysis and the experimental measurements. Whereas
the experimental profiles show a rather narrow front with no
broadening in time, the sedimentation front of the theoretical
profiles gets broader and broader in time. As the local apparent
viscosity $\eta(r)$ of the sheared fluid typically varies by a
factor of 2 in the measurement zone (see
Fig.~\ref{fig_vitesse_visco_locale}b), it is indeed easily
understood that the front corresponding to $\phi=5$\% should move
2 times faster than the front corresponding to $\phi$=0\%, which
is clearly not the case experimentally.

This suggests that collective effects are at play, which tend to
stabilize the front at a given speed basically set by the average
apparent viscosity. This point may not be a surprise. In a
Newtonian fluid, it is actually known that sedimentation
velocities are correlated in the horizontal plane over very long
distances, of order $10 d \phi^{-1/3}$ \cite{Guazzelli2011}. For
the particles we use in our study, this lengthscale is of the
order of 1~cm, i.e. of the order of the gap size (which is
1.9~cm). This may thus explain our observations.

\section{Drag force exerted on a particle in a sheared yield stress fluid: theoretical
analysis}\label{appendix_theory}

We study the translational motion of a single particle of diameter
$d$ embedded in a sheared Herschel-Bulkley fluid. We focus on the
case where translation of the particle is imposed at a constant
velocity $\Vs$ in the direction $\v{e}_3$ orthogonal to the plane
($\v{e}_1$,$\v{e}_2$) of shear. We try to compute the drag force
$\FHB$ exerted by the fluid on the particle as a function of
$\Vs$, of the properties of the particle and of the fluid, and of
the characteristics of shear.

In the studied problem, the particle occupies the domain
$\Omega_p$ with boundary $\partial \Omega_p$ and the fluid
occupies the domain $\Omega_f$ with boundary $\partial \Omega_f =
\partial \Omega_p \cup \partial \Omega_{\infty}$. The fluid domain
is sheared far from the particle at a given overall shear rate
$\Gammadot$. Then the velocity is prescribed on the external fluid
domain boundary $\partial \Omega_{\infty}$
\begin{equation}
  \label{eq:CLinfini}
  \vu =\frac{1}{2}  \Gammadot \left(x_2 \v{e}_1 + x_1 \v{e}_{2}\right)
\end{equation}
The no-slip condition states that at the particle-fluid boundary
$\partial \Omega_p$, the fluid velocity is that of the particle
\begin{equation}
  \label{eq:CLparticule}
  \vu = - \Vs \v{e}_3
\end{equation}
where $\Vs$ denotes the particle velocity ($\Vs$ is positive when
the particle moves downward).

The isotropic tensorial form of the Herschel-Bulkley behavior law
is
\begin{equation}
  \label{eq:LdC-HB}
  \begin{array}{ll}
     \vvtau = 2\left[\left(\tauc + \etaHB
        \gammadot^n\right)/\gammadot\right] \d  & \text{ if }
    \d \neq 0 \\
     1/2\: \vvtau : \vvtau  \le \tauc &  \text{ if }   \d=0
  \end{array}
\end{equation}
with $\d$ the strain rate tensor (i.e. the symmetric part of the velocity gradient) and $\vvtau$ the deviatoric part
of the Cauchy stress tensor. In Eq.~\ref{eq:LdC-HB}, the shear rate
is $\gammadot = \sqrt{2
  \d:\d}$.

The boundary value problem for the motion of a single particle
moving with velocity $\Vs$ in a sheared yield stress fluid is
defined by Eqs.~\ref{eq:CLinfini}, \ref{eq:CLparticule},
\ref{eq:LdC-HB}, the no-body force momentum balance equation
$\dive \cauchy = 0$ where $\cauchy = \vvtau - p \tun$ with $p$ the
pressure, and the incompressibility condition $\dive \vu = 0$.

The drag force exerted by the yield stress fluid on the particle
can be easily computed from the solution of Eqs.~\ref{eq:CLinfini}
to \ref{eq:LdC-HB}
\begin{equation}
  \label{eq:FH}
  \vFHB = \FHB \ez = \int_{\partial \Omega_p} \cauchy \cdot \vn\, dS
\end{equation}
with $\vn$ the outward unit vector normal to the boundary of the
particle.

An analytic solution of this problem is not available, it is thus
not possible to compute $\FHB$ as a function of the material
properties $\tauc$, $\etaHB$ and $n$, the particle diameter $d$
and the loading parameters $\Gammadot$ and $\Vs$. Leaving aside
the possibility to solve numerically this problem, we restrict
in the sequel to the situation of ``slow'' forced motion of the
particle defined by $\Vs \ll \Gammadot d$ (i.e., when the flow
induced by the particle motion can be considered as a secondary
flow as compared to the flow induced by shear). In these
conditions, the solution of the problem defined by
Eqs.~\ref{eq:CLinfini} to \ref{eq:LdC-HB} is looked for as an
asymptotic expansion of the small dimensionless parameter
$\varepsilon = \Vs / \Gammadot d$. It is easily shown that, up to
the second order in $\varepsilon$, the solution of this problem is
\begin{equation}
  \label{eq:dev-vitesse}
\vu = \vuG + \varepsilon \vus + O(\varepsilon^2)
\end{equation}
and
\begin{equation}
  \label{eq:2}
p = \pG + \varepsilon \ps + O(\varepsilon^2)
\end{equation}
where $\vuG$ and $\pG$ are the velocity and pressure in the
sheared yield stress fluid when the particle is at rest, i.e. they
are solutions of Eqs.~\ref{eq:CLinfini} to~\ref{eq:LdC-HB} for
$\Vs=0$, while $\vus$ and $\ps$ are solutions of the boundary
value problem
\begin{equation}
  \label{eq:Pb-ordre1}
  \begin{array}{llll}
  \vus & = &  \vs \ez \quad &(\partial \Omega_p) \\
  \vus & = & 0 \quad  &(\partial \Omega_{\infty}) \\
  \end{array}
 \end{equation}
with $\vs$ defined as $\vs = \Vs / \varepsilon$. The Cauchy stress
tensor is here $\cauchy = \vv{\sigma}_{\text{f}} + \varepsilon
\cauchys + O(\varepsilon^2)$, with
$\vv{\sigma}_{\text{f}}=\vvtauG- \pG \tun$ and $\cauchys = \vvtaus
- \ps \tun$, where $\vvtauG$ and $\vvtaus$ are the zeroth and
first order term of an asymptotic expansion\footnote{Note that
this approach is similar to that used in hydrodynamic stability
analysis, where linearization of the stress field is performed
when adding a perturbation to a base flow \cite{Nouar2007}.}
$\vvtau = \vvtauG + \varepsilon \vvtaus + O(\varepsilon^2)$ of the
constitutive law~\ref{eq:LdC-HB} with respect to $\varepsilon$.
Putting the strain rate tensor associated to the velocity
field~\ref{eq:dev-vitesse} into Eq.~\ref{eq:LdC-HB} yields
\begin{equation}
  \label{eq:taus}
  \vvtaus = 2\left[\left(\tauc + \etaHB
        \gammadotG^n\right)/\gammadotG\right] \ds
        - 2 \frac{\tauc}{\gammadotG} \frac{\dG : \ds}{\dG : \dG} \dG + 2
        \etaHB (n-1) \gammadotG^{n-1} \frac{\dG : \ds}{\dG : \dG} \dG
\end{equation}
where $\dG$ and $\gammadotG = \sqrt{2 \dG: \dG}$ are the strain
rate tensor and the shear rate associated to $\vuG$ while $\ds$ is
the strain rate tensor associated to $\vus$. Eq.~\ref{eq:taus} is
valid only if the fluid deforms ($\gammadotG \neq 0)$ everywhere
in the fluid domain. Such an assumption is undoubtedly fulfilled
here because in the zeroth order problem, the fluid domain is
uniformly sheared on its outside boundary (see
Eq.~\ref{eq:CLinfini}).

When the particle is at rest, the symmetries of the problem
imply that the net force $\int_{\partial \Omega_p}
\vv{\sigma}_{\text{f}}
   \cdot \vn\, dS$ exerted by the fluid on the particle
is zero. When the particle moves slowly (i.e. $\varepsilon \ll
1$), the drag force is thus a first order quantity in
$\varepsilon$
\begin{equation}
  \label{eq:FHB-2}
   \vFHB = \FHB \ez = \varepsilon \int_{\partial \Omega_p} \cauchys
   \cdot \vn\, dS    + O(\varepsilon^2)
\end{equation}

Let us point out that for orthogonal flows, i.e. when $\dG :
\ds=0$, Eq.~\ref{eq:taus} reduces to $\vvtaus = 2\left[\left(\tauc
+ \etaHB \gammadotG^n\right)/\gammadotG\right] \ds$, which
corresponds to a Newtonian behavior, thus justifying the use of
expressions valid for Newtonian materials such as
Eq.~\ref{eq_vit_stokes}. Of course, around a spherical particle,
$\dG : \ds\neq0$ and things are more complex. Because the shear tensor field (Eq.~\ref{eq:taus}) is then a nonlinear function of both the zeroth and first order problem solutions, it is actually not possible to predict even
qualitatively how the force $\FHB$ depends upon quantities
$\Gammadot$ and $\Vs$ in a general case. Nevertheless some results
can be obtained in the `plastic' and `viscous' regimes defined
respectively by $Bi$\m1$\ll 1$ and $Bi$\m1$\gg 1$ where the
inverse Bingham number is
$Bi$\m1$=\eta_{_{H\!B}}\Gammadot^n/\tau_y$.

\subsection*{Plastic regime}

In the `plastic' regime ($\tau_y\gg\eta_{_{H\!B}}\gdot^n$), if $\dG
\neq 0$, the asymptotic expansion of the Herschel-Bulkley law is
\begin{equation}
  \label{eq:LdC-plas}
     \vvtau = 2 \tauc/\gammadotG\:\dG
+   \varepsilon\left(2 \tauc / \gammadotG \ds
        - 2 \frac{\tauc}{\gammadotG} \frac{\dG : \ds}{\dG : \dG} \dG
      \right)
\end{equation}
The leading  term of Eq.~\ref{eq:LdC-plas} corresponds to a rigid
plastic constitutive law, and the zeroth order problem solution is
the velocity field of a perfect rigid-plastic material strained at
a constant shear rate $\Gammadot$ flowing around a particle
rotating freely around a fixed point. If $(\v{v}_{1}, p_1$)
denotes the solution of this problem for $\Gammadot = 1$, a
classic result of the theory of plasticity \cite{Hill1998} asserts that the solution of the problem is $\vuG =
\Gammadot \v{v}_1$, $\pG = p_1$ for any value of $\Gammadot$.

Putting this result into Eq.~\ref{eq:LdC-plas} yields the first
order constitutive law
\begin{equation}
  \label{eq:LdC-plas-1}
  \vvtaus = 2 \tauc / \gammadotG \left(\ds - \frac{\dun : \ds}{\dun :
      \dun} \dun \right)
\end{equation}
with $\dun$ the strain rate tensor associated to $\v{v}_1$.

The first order constitutive law being linear, so is the first
order problem. Thus both $\vvtaus$ and $\ps$ depend linearly upon
$\vs$. It is sufficient to report this property into
Eq.~\ref{eq:FHB-2} to show that the drag force is a linear
function of the quantity $\tauc  / \Gammadot \times \Vs d$. Then
we do have
\begin{equation}
  \label{eq:F-HB-plas}
  \text{If } \quad \frac{\etaHB \Gammadot^n}{\tauc} \ll 1 \quad \text{ and }
  \quad \frac{\Vs}{\Gammadot d } \ll 1  \quad \text{then }\quad  \FHB
\propto \Vs\frac{\tauc}{\Gammadot}  d
\end{equation}

These equations applied in the case of sedimentation,
where the driving force is $\FHB\propto\Delta\rho gd^3$, yield
\bea\Vs\propto\frac{\Delta\rho g d^2}{\tauc/\Gammadot}\eea

\subsection*{Viscous regime}

We now turn to the situation where the viscous stress $\etaHB
\Gammadot^n$ is much larger than the yield stress $\tauc$; the
asymptotic expansion of the Herschel-Bulkley law is then
\begin{equation}
  \label{eq:LdC-visq}
 \vvtau = 2 \etaHB \gammadotG^{n-1} \dG
+ \varepsilon \left(2 \etaHB \gammadotG^{n-1} \ds + 2 (n-1) \etaHB
  \gammadotG^{n-1} \frac{\dG : \ds}{\dG : \dG} \dG \right)
\end{equation}
In this situation, the leading term of Eq.~\ref{eq:LdC-plas}
corresponds to a power law material. Then, it is easily shown that the zeroth order
solution of the problem is $(\vuG = \Gammadot \v{v}_1$, $\pG =
\Gammadot^n p_1)$ for any value of $\Gammadot$ with $(\v{v}_1,
p_1)$ the solution of the first order problem for $\Gammadot = 1$.
Thus, performing similar computations as in
the `plastic' regime, and putting this result into the first order
term of the asymptotic expansion of the constitutive law yields
the linear law
\begin{equation}
\vvtaus = 2 \etaHB \gammadotG^{n-1} \left(\ds + (n-1) \frac{\dun :
    \ds}{\dun : \dun} \dun \right)
\end{equation}
Therefore, we do again obtain a linear first order problem, which
implies that both $\vvtaus$ and $\ps$ depend linearly upon $\Vs
\times \Gammadot^{n-1}$. Finally, in the `viscous' regime the drag
force is given by
\begin{equation}
  \label{eq:F-HB-visq}
\text{If } \quad \frac{\etaHB \Gammadot^n}{\tauc} \gg 1 \quad
\text{ and } \quad \frac{\Vs}{\Gammadot d} \ll 1  \quad \text{then
}\quad \FHB \propto {\Vs}\,\etaHB  {\Gammadot}^{n-1} d
\end{equation}

These equations applied in the case of sedimentation,
where the driving force is $\FHB\propto\Delta\rho gd^3$, yield
\bea\Vs\propto\frac{\Delta\rho g d^2}{\etaHB
{\Gammadot}^{n-1}}\eea

\end{document}